# Bound→free and bound→bound multichannel emission spectra from selectively excited Rydberg states in the ZnAr and CdAr van der Waals complexes

J. Dudek [a], A. Kędziorski [b], J. P. Zobel [c], M. Krośnicki [d], T. Urbańczyk [a], K. Puczka [a], J. Koperski [a]

[a] *Smoluchowski Institute of Physics, Faculty of Physics, Astronomy and Applied Computer Sciences, Jagiellonian University, Łojasiewicza 11, 30-348 Kraków, Poland*

[b] *Institute of Physics, Faculty of Physics, Astronomy and Informatics, Nicolaus Copernicus University, Grudziądzka 5/7, 87-100 Toruń, Poland*

[c] *Institute of Theoretical Chemistry, Faculty of Chemistry, University of Vienna; Währinger Straße 17, A-1090 Vienna, Austria*

[d] *Institute of Theoretical Physics and Astrophysics, Faculty of Mathematics, Physics and Informatics, University of Gdansk, Wita Stwosza 57, 80-308 Gdańsk, Poland*

**Abstract**

Multichannel dispersed emission spectra recorded upon a selective excitation of Rydberg electronic energy states in the ZnAr and CdAr van der Waals (vdW) complexes are analysed as a proof-of-concept of the future experimental approach. Simulations of the emission spectra are based on *ab-initio* calculated interatomic potentials and transition dipole moments (TDMs). Experimental set-up that is under construction along with the experimental procedure are discussed.

**1. Introduction**

Bound→free and bound→bound laser induced fluorescence (LIF) emission (also called fluorescence) spectra emitted after a selective excitation of Rydberg states of diatomic van der Waals (vdW) complexes are a source of information, mainly on repulsive part, but also on potential well of the lower-lying electronic state potentials that are not accessible in the excitation from the ground state. Modelling of the dispersed emission spectra provides a direct information on the steepness of the short-range branch of the potential at which the emission terminates – information that is hard to access from LIF excitation spectra.

The selective excitation to the Rydberg states can be realized using different methods. If the transition from the ground to the Rydberg state is allowed, a direct excitation can be implemented. This way, for complexes such as *e.g.*, the $Rg_2$ (Rg=rare gas atom), pulsed synchrotron radiation has been used to perform a state-selective excitation of the $Ar_2$, $Kr_2$ and $Xe_2$ complexes to their $^3 0_u^+(^3P_1)$ or $^3 1_u(^3P_1)$ lower-lying excited states at 106.8-108.8 nm, 124-126.5 nm and 148-151.5 nm wavelength ranges, respectively [1]. Following the excitation, oscillatory bound→free continua, also called Condon internal diffraction (CID) patterns [2], were observed using time-resolved emission in 130-180 nm, 150-200 nm and 175-230 nm spectral ranges, respectively,.

Beside high-energy UV synchrotron radiation beam, a more convenient method for exciting molecular Rydberg states is optical-optical double resonance (OODR) method that simply uses two laser beams, both in the VIS and/or UV range. Also, OODR allows to excite a molecule to the electronic energy states to which transitions from the ground state are forbidden as well as using different intermediate electronic states to access and probe different parts of the ro-vibrational energy structure in the final Rydberg-state potential. Employing OODR, a number of Rydberg electronic states in the MRg (M=Cd, Hg) heteroatomic complexes has been reached, among them, as first, the $C^31(7^3S_1)$ double-well Rydberg state in the HgAr that was excited *via* the $A^30^+(6^3P_1)$ or $B^31(6^3P_1)$ intermediate states [3,4], and studied recording LIF excitation spectra. The same intermediates were used to excite to and record excitation spectra from the triplet Rydberg series of the $^3\Sigma^+(n^3S_1, n = 7 - 10)$ states in HgNe [5,6], the $^3\Sigma^+(8^3S_1)$ Rydberg state in HgAr [6] and the singlet Rydberg series of the $^1\Sigma^+(n^1S_0, n = 7 - 9)$ states in HgNe [7].

As far as the CdRg (Rg=Ne, Ar, Kr) complexes are concerned, OODR *via* the lower-lying $A^30^+(5^3P_1)$ and $B^31(5^3P_1)$ intermediates was employed to investigate a complex double-well structure of the $E^31(6^3S_1)$ Rydberg state in CdNe [8], CdAr [9-15] and CdKr [12,14,16,17] using LIF excitation spectra.

Regarding the ZnRg complexes, only the $E^3\Sigma^+(5^3S_1)$ state in ZnAr has been investigated, however, using excitation *via* vaporization-optical (VO) method [18]. In general, VO relies on utilizing a vaporization laser pulse in the first-step of the excitation. It produces a number of atoms in long-lived metastable intermediate states. Characterisation of the $E^3\Sigma^+$ state has been performed using LIF excitation spectra recorded using the second-step $E^3\Sigma^+ \leftarrow a^3\Pi_{0^-}(4^3P_0), b^3\Pi_2(4^3P_2)$ transitions.

As concerns the M$_2$ (M=Zn, Cd, Hg) homoatomic complexes, by employing OODR it was possible for the first time to excite a *gerade* Rydberg state in Cd$_2$ *via* the $b^30_u^+(5^3P_1)$ intermediate and record a pronounced $v' \leftarrow v'' = 0,3$ progressions showing their isotopologue structure [19].

It has to be emphasized however, that in any of the above Rydberg-state studies emission spectra in MRg complexes have not been observed, except one investigation. For the HgAr, Duval *et al.* [4] recorded dispersed emission bound→bound spectra using the $C^31(7^3S_1), v' = 3 \rightarrow A^30^+(6^3P_1), v'', C^31, v' = 14,19 \rightarrow a^30^-(6^3P_0), v''$ and $C^31, v' = 2,4,14 \rightarrow b^32(6^3P_2), c^31(6^3P_2), d^30^-(6^3P_2), v''$ transitions, as well as long bound→free undulated CID *reflection* [20] patterns using the $C^31, v' = 3, 11 \rightarrow a^30^-, C^31, v' = 2, 10 \rightarrow A^30^+, B^31$ and $C^31, v' = 3, 14 \rightarrow b^32, c^31, d^30^-$ transitions; it allowed to probe and determine lower-lying state potentials including their repulsive branches.

In this article, we propose a scheme for excitation of the ZnAr complex to the $^11(4^1D_2)$ Rydberg state *via* the $C^11(4^1P_1)$ intermediate using OODR and detection of a dispersed emission from the $^11(4^1D_2)$ state to the lower-lying electronic states (see Fig. 1a). The goal is to analyse the $^11(4^1D_2) \rightarrow a^30^-(4^3P_0), A^30^+(4^3P_1), B^31(4^3P_1), b^32(4^3P_2), c^31(4^3P_2), d^30^-(4^3P_2), C^11(4^1P_1), D^10^+(4^1P_1)$ emission channels taking into account newly calculated *ab* *initio* ZnAr interatomic potentials and transition dipole moments (TDMs) for transitions between the emitting and lower electronic energy states.

Along with the analysis for the ZnAr, we present also a similar consideration for detection of dispersed emission spectra in the CdAr complex after a selective excitation of the $E^31(6^3S_1)$ Rydberg state (see Fig. 2). The analysis is based on *ab-initio* calculated interatomic potentials obtained in our laboratory and recently published by Krośnicki *et al.* [21].

## 2. *Ab-initio* calculations

Ab initio potentials of electronic states of ZnAr complex up to Zn $6^1S_0$ asymptote have been calculated within three-step approach. Firstly, the state-average complete active space self-consistent field method has been applied (SA-CASSCF) [22,23] for the active space, in which two electrons are distributed on the molecular orbitals of the predominant Zn 4s, 4p, 4d, 5s, 5p, 6s character. Secondly, the multi-state second-order CAS perturbation theory method (MS-CASPT2) has been used [24]. Finally, the spin-orbit interaction has been taken into account by means of restricted active space state interaction method (RASSI-SO) [25]. Scalar relativistic effects have been accounted for within Douglas-Kroll-Hess second-order Hamiltonian [26,27]. We used new ANO-type (ANO=atomic natural orbitals) basis for Zn atom within general contraction scheme (23s17p12d6f4g2h)/[11s9p7d4f2g], which is modified and extended Zn basis from Ref. [28]; discussion on the construction of new Zn ANO basis set can be found in Supporting Information of Ref. [29]. For Ar, we used (17s12p5d4f2g)/[8s7p5d4f2g] ANO basis set [30]. Additionally, we used mid-bond basis applying (8s4p3d1f)/[6s4p3d1f] Hydrogen ANO basis [31]. Details of new Zn basis and of the computational aspects are presented elsewhere [29,32]. TDMs have been calculated within SA-CASSCF/MS-CASPT2/RASSI-SO approach, where the initial and final states are the linear combinations of the SA-CASSCF states. All the calculations were performed with Molcas 8.2 [33] within $C_{2v}$ symmetry.

Details of *ab initio* calculations of electronic states and TDMs of CdAr complex are presented in Ref. [21].

Figure 1. (a) Ab initio interatomic potentials for electronic states of the ZnAr used in simulation of multichannel emission from the $^11(4^1D_2), v' = 13$. The $^11(4^1D_2), v' = 13 \leftarrow C^11, v'' = 10 \leftarrow X^10^+, v = 0$ excitation in OODR process is depicted with vertical solid arrows whereas allowed emission channels are shown with vertical dashed arrows. (b) Ab initio interatomic potentials intersecting with the $^11(4^1D_2)$ emitting state (only these with Ω=1 are shown). (c), (d), (e) Values of |TDM|² for an emission from the $^11(4^1D_2)$ state to the states correlating with the $4^1P_1$, $4^3P_J$, and $4^1S_0$ atomic asymptotes. (f), (g) Values of |TDM|² for the $C^11(4^1P_1) \leftarrow X^10^+(4^1S_0)$ excitation (thick solid line) and the excitation from the $C^11$ (thin solid and dotted lines) to several Rydberg states in ZnAr.

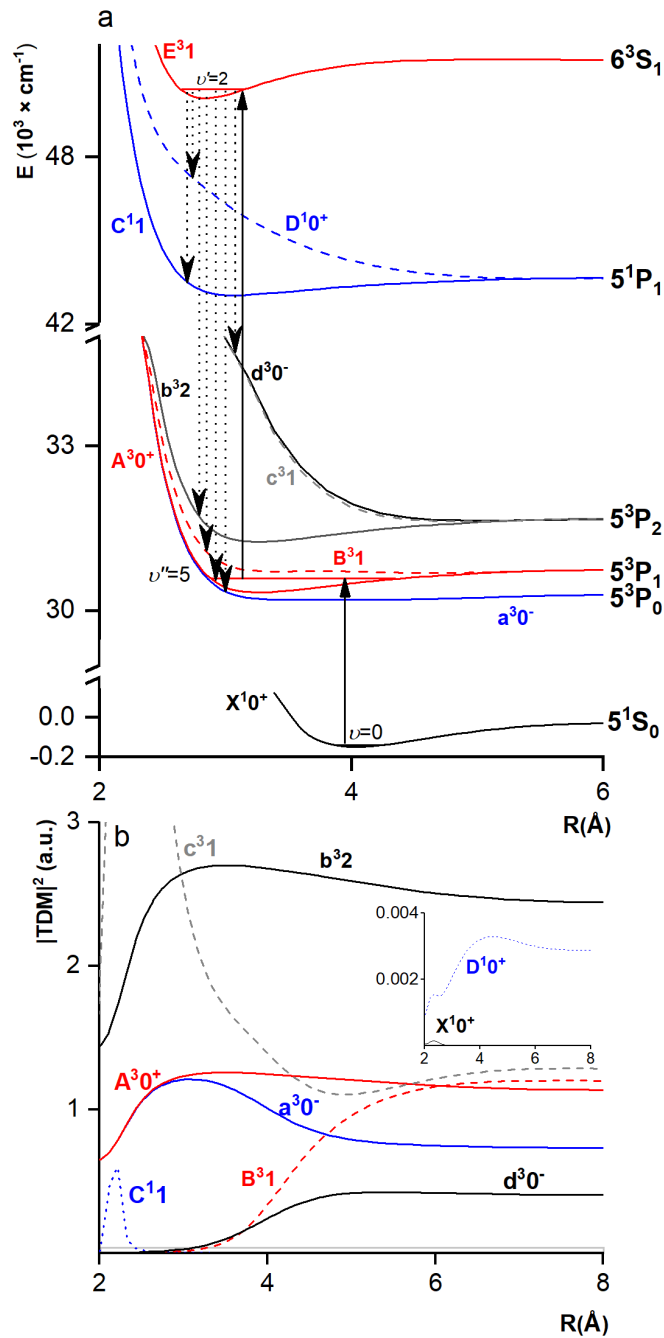

Figure 2. (a) Ab initio interatomic potentials for electronic states of the CdAr [21] used in simulation of multichannel emission from the $E^31(6^3S_1), v'=2$. The $E^31, v'=2 \leftarrow A^30^+(5^3P_1), v''=5 \leftarrow X^10^+(5^1S_0), v=0$ excitation in OODR process is depicted with vertical solid arrows whereas allowed emission channels are shown with vertical dashed arrows. (b) Values of |TDM|$^2$ for emission from the $E^31$ state to the states correlating with the $5^1P_1$, $5^3P_J$ and $5^1S_0$ atomic asymptotes. Inset shows two smallest |TDM|$^2$.

## 3. Experiment – a prospective set-up

There are two experimental methods devoted to detection of LIF dispersed emission spectra with the use of: i) a tuneable monochromator in front of a photocathode of a photomultiplier tube (PMT) - the method which is limited by a considerable time consumption during the experiment, especially when a large spectral range is scanned over (see *e.g.*, [36]), or ii) a spectrometer fitted with an iCCD camera (see *e.g.*, [37]) - the method in which the whole spectrum is recorded at once without a necessity of executing the time-consuming scans.

Figure 3 presents experimental set-up for excitation of the ZnAr to the selected Rydberg state using OODR method, followed by detection of a dispersed emission that occurs to the lower-lying electronic energy states. Description of a continuously operating source-module of a molecular beam containing zinc and description of a molecular single excitation were included elsewhere [34]. Both crucible and cap with the source nozzle are made of graphite that is chemically resistant to numerous melted aggressive metals and metal vapours. ZnAr complexes in the beam are excited by UV laser radiation produced by a pulsed dye laser (TDL90, Quantel) pumped with a Nd$^+$:YAG laser (YG981C, Quantel). To perform OODR excitation, a second dye laser (TDL90, Quantel) pumped with another Nd$^+$:YAG laser (Powerlite 7010, Continuum) is employed. Total LIF emission from the first and the second excitation-step is controlled by two photomultipliers (R375, Hamamatsu and 9893QB/350, Electron Tubes) and recorded with a digital oscilloscope (TDS 2024B, Tektronix). Emission from the Rydberg state is focused on the entrance slit of an echelle spectrometer (ME5000, Andor), dispersed with a spectral resolution $\frac{\lambda}{\Delta\lambda} = 5000$ and operating wavelength range 200-975 nm, and detected by iCCD camera (334T, iStar). The lasers, oscilloscope and camera are triggered by a digital delay generator (DDG) (DG645, Stanford Research Systems).

The experimental set-up for excitation of vdW complexes containing cadmium atoms using OODR method was used in our previous studies [14]. In the experiment, a molecular beam was produced with high-temperature, high-pressure pulsed molecular beam-module [35]. For the CdAr, as in the case of the ZnAr, spectrometer fitted with iCCD camera triggered with DDG is planned to be installed to record emission from the Rydberg state.

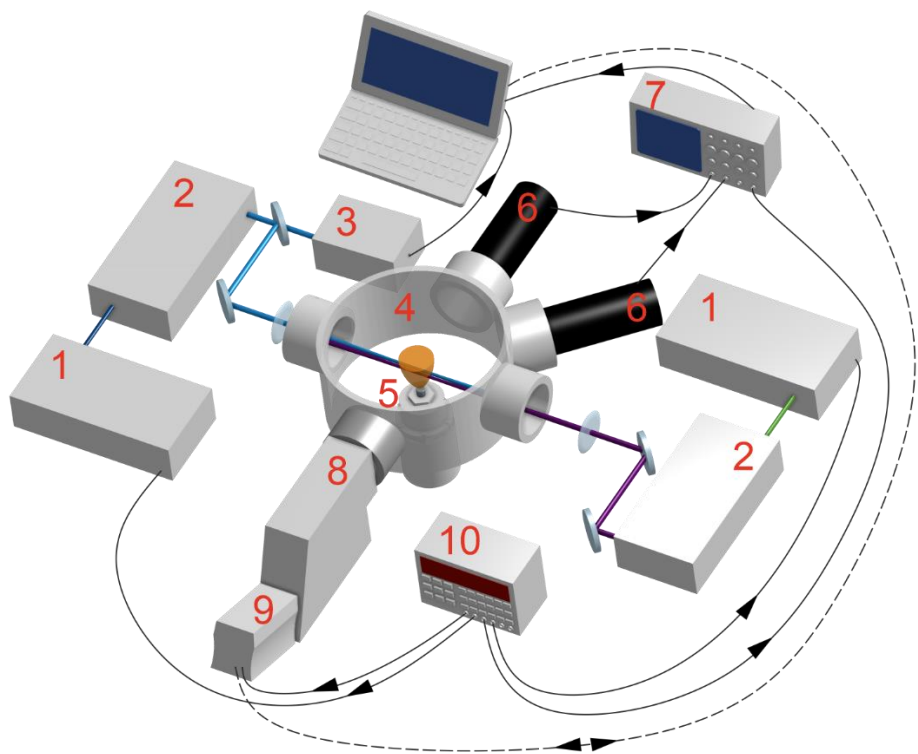

Figure 3. Scheme of experimental set-up for excitation of the ZnAr complexes to the $^1 1(4^1D_2)$ Rydberg state using OODR method followed by detection of a dispersed emission to lower-lying electronic energy states (under construction in our laboratory). 1 - Pulsed Nd$^+$:YAG lasers, 2 - dye lasers, 3 - wavemeter, 4. - vacuum chamber, 5 - continuously operating source-module of the molecular beam, 6 – photomultipliers, 7- oscilloscope, 8 - spectrometer, 9 - iCCD camera, 10 - digital delay generator.

## 4. Results

For both ZnAr and CdAr complexes, LEVEL Fortran code [38] for vibrational spectra, and PGOPHER [39] for rotational spectra, were used to simulate all bound↔bound transitions in OODR excitations and subsequent emissions. To simulate bound→free CID patterns in emission, BCONT program [40] was used. The initial states (the $^1 1(4^1D_2)$ and $E^3 1(6^3S_1)$ in ZnAr and CdAr, respectively) from which emission was originated were selected to assure the allowed emission channels. Indeed, the projection of the total electronic angular momentum represented by $\Omega$ quantum number (Hund's case c) equals 1 for both the $^1 1(4^1D_2)$ and $E^3 1(6^3S_1)$ emitting states, and thus transitions to the lower-lying states with $\Omega$ = 0,1 and 2 are allowed.

*4.1. ZnAr*

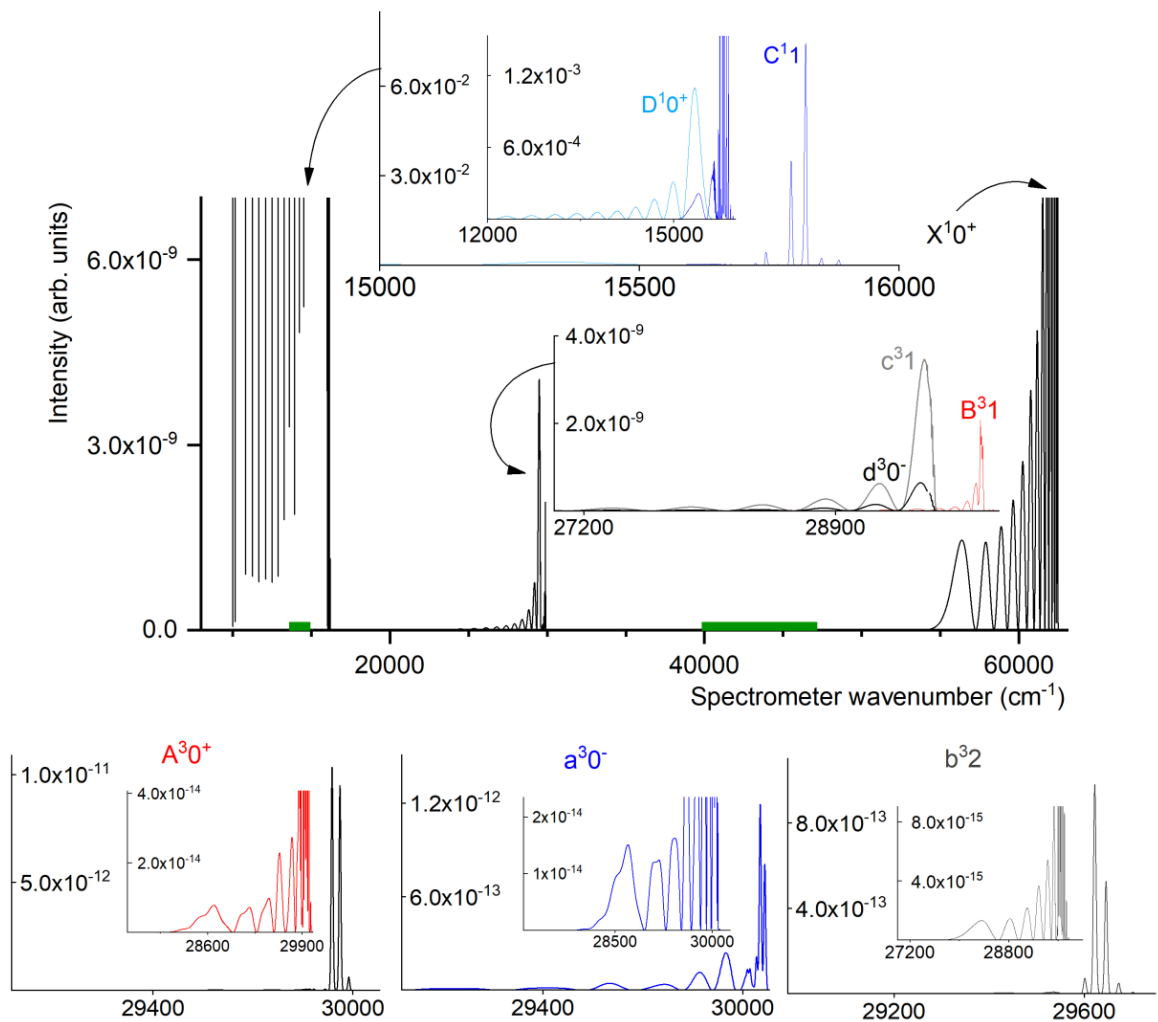

Figure 4. Simulation [38,39,40] of multichannel bound→bound and bound→free transitions in emission from the $^1 1(4^1D_2), v' = 13$ in ZnAr to the states correlating with the $4^1P_1$, $4^3P_J$ and $4^1S_0$ atomic asymptotes. Bound→bound transitions include vibrational and rotational spectra, Gaussian convolution (FWHM) is equal to spectrometer resolution, *i.e.* $\frac{\lambda}{\Delta\lambda} = 5000$. Insets show individual single-channel bound→bound and bound→free transitions. Two green bars on energy axis show a location of the expected emission spectrum from the $C^1 1, v' = 10$ intermediate level used in OODR excitation (based on [43] and simulation). See text for details.

Analysis of |TDM(R)|² functions for the excitation from the ground $X^1 0^+, v=0$ reveals that the excitation to the $C^1 1(4^1P_1)$ state can be chosen for being the intermediate step in OODR. Experimental spectrum of the $C^1 1, v'' \leftarrow X^1 0^+, v = 0$ transition [41, 42] reveals the $v'' = 10 \leftarrow v = 0$ as the most intensive vibrational transition. Therefore, the $C^1 1, v'' = 10$ was chosen for first-step excitation in OODR. In the previous realization of the $C^1 1 \leftarrow$

$X^1 0^+$ excitation [41], frequency-doubled output of a dye laser working with Stilbene 3 dye was used, but a frequency-tripling with DCM dye can be applied as well. Due to wide-separation of the isotopologue components in the $v'' = 10 \leftarrow v = 0$ transition, an isotopologue-selective OODR [14] is possible to apply. Thus, the $^1 1(4^1 D_2) \leftarrow C^1 1, v'' = 10$ can be realized mostly for the $^{64}Zn^{40}Ar$ isotopologue.

For the final state in OODR, the $^1 1(4^1 D_2)$ electronic state was chosen not only to assure all transition channels in the subsequent emission to be allowed, but also to assure the excitation frequencies from dye lasers available in our lab. Realization of the $^1 1(4^1 D_2), v' \leftarrow C^1 1, v'' = 10$ transition requires frequencies in the range of 15200-15850 cm$^{-1}$ (it is ensured by using *e.g.*, DCM dye pumped with a second harmonic of Nd$^+$:YAG laser). Although several intersecting potentials with Ω=1 are accessible in the energy region: $^1 1(4^1 D_2)$, $^3 1(4^3 D_1)$, $^3 1(4^3 D_2)$, $^3 1(4^3 D_3)$ and $^1 1(5^1 P_1)$ (see Fig. 1b), the $^1 1(4^1 D_2) \leftarrow C^1 1(4^1 P_1), v'' = 10$ transition has the highest intensity. It is due to the high value of |TDM|$^2$ (see Fig. 1f) and the same multiplicity of the initial and final states. Simulation of the $^1 1(4^1 D_2), v' \leftarrow C^1 1, v'' = 10$ for the $^{64}Zn^{40}Ar$ isotopologue (Fig. 5) includes rotational transitions with the rotational temperature $T_{rot}$=10 K and Gaussian convolution FWHM = 0.2 cm$^{-1}$, that are typical values for excitation experiments realized in our laboratory. The strongest $v' = 13$ vibrational component was chosen to simulate emission from.

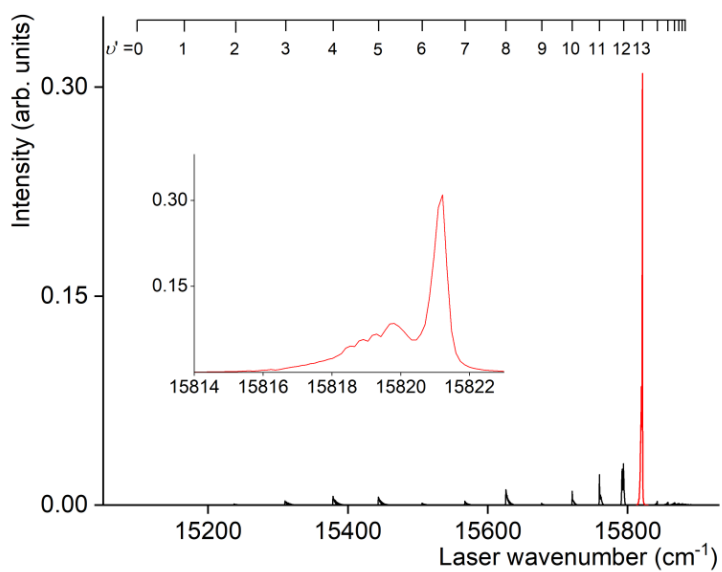

Figure 5. Simulation of the $^1 1(4^1 D_2), v' \leftarrow C^1 1(4^1 P_1), v'' = 10$ transition in excitation spectrum of ZnAr that is planned for realization of OODR scheme. The simulation was performed for the $^{64}Zn^{40}Ar$ isotopologue, with parameters: T$_{rot}$ = 10 K and 0.2 cm$^{-1}$ (FWHM) Gaussian convolution. Inset shows the spectrum of the $v' = 13 \leftarrow v'' = 10$ transition with partly resolved rotational structure.

In the emission from the $^1 1(4^1 D_2), v' = 13$, parameters used in the simulation of bound→bound transitions correspond to the conditions that are most likely to be realized in the planned experiment, *i.e.* rotational temperature $T_{rot}$ = 10 K and spectral broadening of bound→bound components that corresponds to the spectrometer resolution ($\frac{\lambda}{\Delta\lambda}$ = 5000).

The total multichannel emission (bound→free along with bound→bound transitions) from the $^1 1(4^1D_2), v' = 13$ (see Fig. 4) encompasses three main energy regions: (i) 10000-15900 cm$^{-1}$ where relatively intensive emission to the states correlating with the $4^1P_1$ atomic asymptote is expected; (ii) 27100-30000 cm$^{-1}$ with much weaker spin-forbidden emission to triplet states correlating with the $4^3P_J$ asymptotes; (iii) emission to the ground $X^1 0^+$ state that is beyond of the range of the spectrometer (see Sec. 3).

According to the simulation (Fig. 4), there is a possibility of detection of strong bound→free emission to the $D^1 0^+(4^1P_1)$ state which within the 12100-15100 cm$^{-1}$ energy range is not influenced by other emission channels. Only the first maximum within the 15100-15600 cm$^{-1}$ overlaps with residual bound→free emission from the $^1 1(4^1D_2), v' = 13$ emitting state.

Bound→free transitions to the $c^3 1(4^3P_2)$ and $d^3 0^-(4^3P_2)$ states overlap in the energy region 29600-32300 cm$^{-1}$. A higher value of |TDM|$^2$ for the $^1 1(4^1D_2) \rightarrow c^3 1$ transition below R = 3 Å results in a considerably higher intensity of the emission to the $c^3 1$ state.

Within the 29600-29900 cm$^{-1}$ range, a combination of bound→bound and bound→free transitions to the $a^3 0^-(4^3P_0)$, $A^3 0^+(4^3P_1)$ and $B^3 1(4^3P_1)$ states is located, with a significant domination of these to the $B^3 1$ state. Although resolution of the spectrometer is too low to resolve the bound→bound $^1 1(4^1D_2) \rightarrow B^3 1$ emission spectrum (below 1 cm$^{-1}$ is required), the bound→free continuum could be detectable.

It has to be noted, that the emission from the intermediate $C^1 1, v' = 10$ used in OODR to the states correlating with the $4^3P_J$ atomic asymptotes occurs within the 13500-14000 cm$^{-1}$ range, so, it could overlap with the bound→free $^1 1(4^1D_2), v' = 13 \rightarrow D^1 0^+$ transition. However, change of the multiplicity in former transitions suppress the intensity of their spectra. According to Koperski and Czajkowski [43], strong $C^1 1, v'' = 10 \rightarrow X^1 0^+$ emission occurs within the 39900-47000 cm$^{-1}$ range, therefore, does not overlap with any emission channels from the $^1 1(4^1D_2)$ (see green bar in Fig. 4).

*4.2. CdAr*

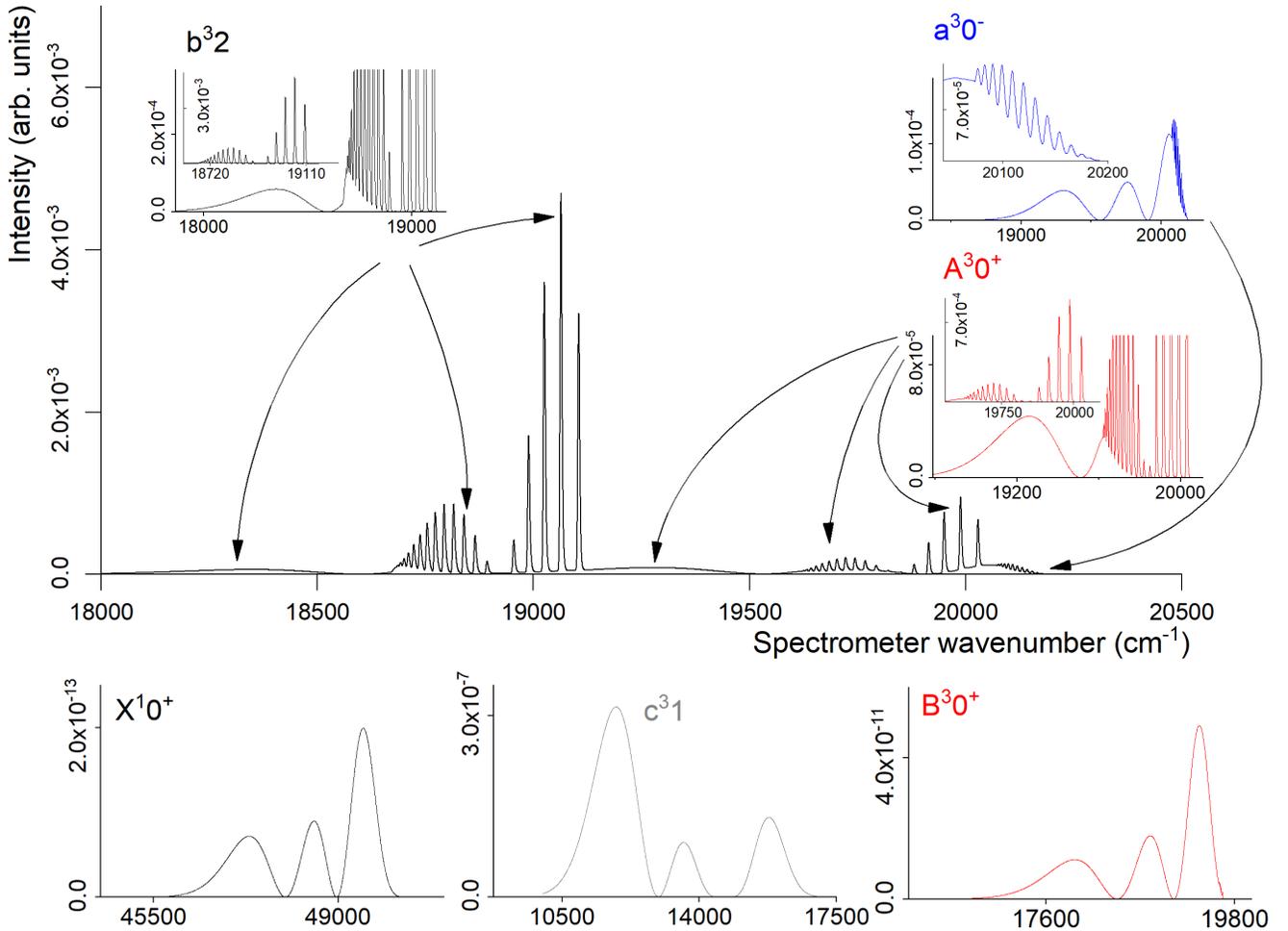

Figure 6. Simulation of multichannel bound→bound and bound→free transitions in emission from the $E^31(6^3S_1), v' = 2$ in CdAr to the states correlating with the $5^3P_J$ and $5^1S_0$ atomic asymptotes. Bound→bound transitions include vibrational and rotational spectra, Gaussian convolution (FWHM) is equal to spectrometer resolution, *i.e.* $\frac{\lambda}{\Delta\lambda} = 5000$. Individual single-channel bound→bound and bound→free components are shown in insets. Low-intensity spectra are shown below the main plot. See text for details.

In case of the CdAr, excitation from the ground to the double-well $E^31(6^3S_1)$ Rydberg state using OODR scheme, through $v'' = 5$ level in the $A^30^+(5^3P_1)$ intermediate state, has been recently recorded and re-analysed [15]. To simulate emission from the $E^31$ state, $v' = 2$, situated near the bottom of the $E^31$-state inner well, was chosen as the $E^31, v' = 2 \leftarrow A^30^+, v'' = 5$ transition possess a relatively large intensity. It also allowed to avoid an influence of the $E^31$-state outer-well vibrational levels on those in the inner-well as far as calculation of their wave functions and energies was concerned.

In the emission from the $E^31, v' = 2$, $T_{rot}$ and spectral broadening used in the simulation of bound→bound transitions are the same as in case of the ZnAr analysed above. For the CdAr, the most intense emission spectra extend within the 18000-20200 cm$^{-1}$ range. According to the simulation, the most intensive $E^31, v' = 2 \rightarrow b^32(5^3P_2)$, $a^30^-(5^3P_0)$ and $A^30^+(5^3P_1)$ channels are generally well separated. Only the bound→free transitions in emission to the $a^30^-$ and bound→bound transitions to the $A^30^+$ states overlap. Taking into account the experimental determination of the $A^30^+$-state potential that was based on the $A^30^+, v'' \leftarrow X^10^+, v = 0$ vibrational structure [15] and that the overlapped transitions represent different character (discrete and continuous), both spectra would be able to separate. However, a particularly valuable observation from the simulation of the emission spectra, that would extend possible characterization of CdAr electronic energy states, is well-resolved vibrational structure in the $b^32$ and $a^30^-$ states, as their excitation from the ground state is forbidden.

## 5. Conclusions

Multichannel emission spectra from $v' = 13$ level in the $^11(4^1D_2)$ and $v' = 2$ level in the $E^31(5^3S_1)$ Rydberg states in ZnAr and CdAr, respectively, were simulated based on *ab initio*-calculated potentials and transition dipole moment (TDM) functions for ZnAr [32] and CdAr [21]. An experimental realization of the detection of the emission spectra presently prepared in our laboratory was proposed and discussed. Because an excitation of the $^11(4^1D_2)$ Rydberg state in ZnAr has never been reported, OODR excitation via the previously reported [41,42] $C^11(4^1P_1)$ intermediate was simulated as well.

Simulation of emission spectra from the $^11(4^1D_2), v' = 13$ in ZnAr reveals a promising prospect for recording bound→free transitions to the $D^10^+(4^1P_1)$, $c^31(4^3P_2)$ and $B^31(4^3P_1)$ states that would allow determine a shape of short-range repulsive parts of the potentials at which emission terminates, and/or explore potentials of lower-lying electronic states that are not accessible in the excitation from the ground state.

In case of emission from the $E^31, v' = 2$ state in CdAr, according to the simulation, bound→bound and bound→free transitions to the $b^32(5^3P_2)$, $a^30^-(5^3P_0)$ and $A^30^+(5^3P_1)$ are well–separated or easy to be separated in detection. It offers prospects for determination of the shape of both repulsive part and the potential well in one experiment.

**Acknowledgements**

The project was financed by the National Science Centre Poland under Contract No. UMO-2015/17/B/ST4/04016. A partial support from Polish Ministry of Science and Higher Education through grants 7150/E-338/M/2018 and 7150/E-338/M/2019 is acknowledged.